\documentclass{webofc}
\usepackage[varg]{txfonts}   % Web of Conferences font
\newcommand{\pp}           {pp\xspace}
\newcommand{\PbPb}         {\mbox{Pb--Pb}\xspace}

\newcommand{\snn}          {\ensuremath{\sqrt{s_{\mathrm{NN}}}}\xspace}
\newcommand{\pt}           {\ensuremath{p_{\rm T}}\xspace}

\newcommand{\Raa}         {\ensuremath{R_{\rm AA}}\xspace}

\newcommand{\dEdx}         {\ensuremath{\textrm{d}E/\textrm{d}x}\xspace}

\newcommand{\twosevensix}  {$\sqrt{s}~=~2.76$~Te\kern-.1emV\xspace}
\newcommand{\five}         {$\sqrt{s}~=~5.02$~Te\kern-.1emV\xspace}
\newcommand{\twosevensixnn}{$\sqrt{s_{\mathrm{NN}}}~=~2.76$~Te\kern-.1emV\xspace}
\newcommand{\fivenn}       {$\sqrt{s_{\mathrm{NN}}}~=~5.02$~Te\kern-.1emV\xspace}
\newcommand{\GeVc}         {Ge\kern-.1emV/$c$\xspace}
\newcommand{\MeVc}         {Me\kern-.1emV/$c$\xspace}
\newcommand{\TeV}          {Te\kern-.1emV\xspace}

\newcommand{\jpsi}         {\ensuremath{\text{J}/\psi}\xspace}

\begin{document}
\title{Quarkonia as probes of the QGP and of the initial stages of the heavy-ion collision with ALICE}

\author{\firstname{Ingrid} \lastname{McKibben Lofnes}\inst{1}\fnsep\thanks{\email{ingrid.mckibben.lofnes@cern.ch}}, for the ALICE Collaboration
}

%\author{\firstname{Ingrid McKibben Lofnes}\inst{1}\lastname{, on behalf of the ALICE Collaboration}\fnsep\thanks{\email{ingrid.mckibben.lofnes@cern.ch}} 
        %\firstname{Second author} \lastname{Second author}\inst{2}\fnsep\thanks{\email{Mail address for second
        %     author if necessary}} \and
        %\firstname{Third author} \lastname{Third author}\inst{3}\fnsep\thanks{\email{Mail address for last
        %     author if necessary}}
        % etc.
%}

\institute{University of Bergen (Norway)
%\and
%           the second here 
%\and
%           Last address
          }

\abstract{%
%  Insert your english abstract here.
Studies of quarkonium production in heavy-ion collisions can be used for probing QGP properties. The suppression and regeneration of bound quarkonium states is sensitive to the medium properties. Modifications of the quarkonium polarization in \PbPb collisions with respect to \pp collisions may give further insight into the suppression and regeneration mechanisms in the QGP. Quarkonia are also sensitive to the initial stages of heavy-ion collisions, and measurements in photonuclear collisions may help constrain the nuclear gluon-distribution at low Bjorken-$x$.
In this work, recent quarkonium measurements performed by ALICE in \PbPb collisions at \fivenn will be discussed. Preliminary measurements of the inclusive \jpsi \Raa measured at both forward and midrapidity will be presented. The \jpsi  polarization measured for the first time in \PbPb collisions, as well as preliminary measurements of the coherent \jpsi photoproduction cross section, will be discussed. 
}
\maketitle
%
%\section{Introduction}
%\label{intro} 
The primary goal of the ALICE experiment is the study of strongly-interacting matter, known as the quark--gluon plasma (QGP), expected to be created during high-energetic heavy-ion collisions. Quarkonium production is sensitive to the medium produced in these collisions and serves as a probe of the QGP properties. Since the heavy quarks (i.e. charm and beauty) are produced in the initial hard partonic scattering, they experience the full evolution of the fireball. Bound quarkonium states immersed in a strongly interacting medium may dissociate, suppressing the quarkonium production compared to binary-scaled proton--proton (\pp) collisions at the same energy. This suppression is actually a subtle interplay of several mechanisms, among them the screening of free color charges in the medium~\cite{Matsui:1986dk}. Recent theoretical developments have shown that the melting of quarkonium states is a dynamical process, where the binding of quarkonium states is weakened over time. This weakening depends on how the medium interacts with the bound states and on the time spent in the medium~\cite{Rothkopf:2019ipj}. 
If the abundance of heavy quark pairs is large enough, a regeneration process may take place, where heavy quarks (re)combine either at the QGP phase boundary~\cite{Braun-Munzinger:2000csl} or throughout the evolution of the fireball~\cite{coalescence:2001}. 
Recent ALICE measurements are consistent with (re)generation of charm quarks as a dominant source of the \jpsi production at low transverse momentum (\pt)~\cite{ALICE:midRap, ALICE:forwardRap}. %However, the question as to whether the (re)combination takes place solely at the phase boundary or throughout the QGP phase remains unsolved. 
%ALICE measurements of the bottomonium production show strong suppression and are consistent with theoretical predictions with little or no contribution from (re)generation, as the beauty cross section is expected to be much smaller than the charm cross section~\cite{ALICE:Upsilon}.
%If heavy-flavour quarks thermalize in the QGP, one expects the regenerated quarkonium states to participate in the collective motion of the bulk matter. ALICE measurements of the \jpsi elliptic and triangular flow show positive values at both forward and midrapidity~\cite{ALICE:flow}, while the elliptic flow of the $\Upsilon (1\text{S})$is consistent with zero~\cite{upsilonV2}.
%When comparing the \jpsi elliptic flow to that of light hadrons and D mesons, a clear mass hierarchy is observed at low \pt. At high \pt the measured flow seems to converge indicating that the anisotropy in this kinematic region is dominated by path-length dependent energy-loss effects. A similar mass hierarchy is observed for the triangular flow, supporting the hypothesis of charm quarks being kinetically equilibrated in the QGP medium. The \jpsi elliptic flow shows a strong dependence on the centrality, as the flow grows from the most central collisions to cemi-central collisions. The \jpsi triangular flow shows no significant dependency on the centrality.
%The \jpsi elliptic flow is well described by transport models at low \pt but the models underpredict the measured flow above 4 \GeVc. 
A modification of the quarkonium vector state polarization measured in \PbPb collisions with respect to \pp collisions can shed light on quarkonium suppression and regeneration mechanisms in the QGP. 
In addition, quarkonia are sensitive to the initial state of the heavy-ion collisions. Looking at photoproduction of \jpsi mesons in heavy-ion collisions at the LHC, one may constrain the nuclear gluon distribution at low Bjorken-$x$, in the range $x \sim 10^{-5} - 10^{-2}$~\cite{gluonShadowing}. The coherent photoproduction of \jpsi mesons, which takes place at very low \pt ($\sim 60$ \MeVc),  is characterized by the coherent coupling of a photon emitted by one of the nuclei while the target nucleus remains intact. Coherent photoproduction is studied in ultra-peripheral collisions, however recent measurements from peripheral \PbPb collisions show an excess yield of \jpsi at very low \pt, assumed to originate from coherent \jpsi photoproduction~\cite{ALICE:excessYield}.
In these proceedings a selection of recent quarkonium measurements obtained by ALICE at both forward and central rapidity will be discussed. 
\\
\\
\indent  The ALICE detector is capable of measuring quarkonium states down to $\pt = 0$ at both forward and central rapidity. At forward rapidity ($2.5 < \textit{y} < 4$) inclusive quarkonium states are measured through the dimuon channel using the muon spectrometer. The muon spectrometer includes a front absorber for filtering muons, five tracking stations, a dipole magnet with a 3 Tm field integral, two trigger stations and an iron wall to reject punch-through hadrons and low momentum muons. 
At midrapidity ($|\textit{y}| < 0.9$), charmonium states are reconstructed through the dielectron channel using the Inner Tracking System (ITS) composed of six cylindrical layers of silicon detectors and the Time Projection Chamber (TPC) providing track reconstruction and particle identification via the measurement of specific energy loss, \dEdx. A more detailed description of the ALICE detectors can be found in~\cite{Jinst}.

\begin{figure}[h]
\centering
\includegraphics[width=8.cm,clip]{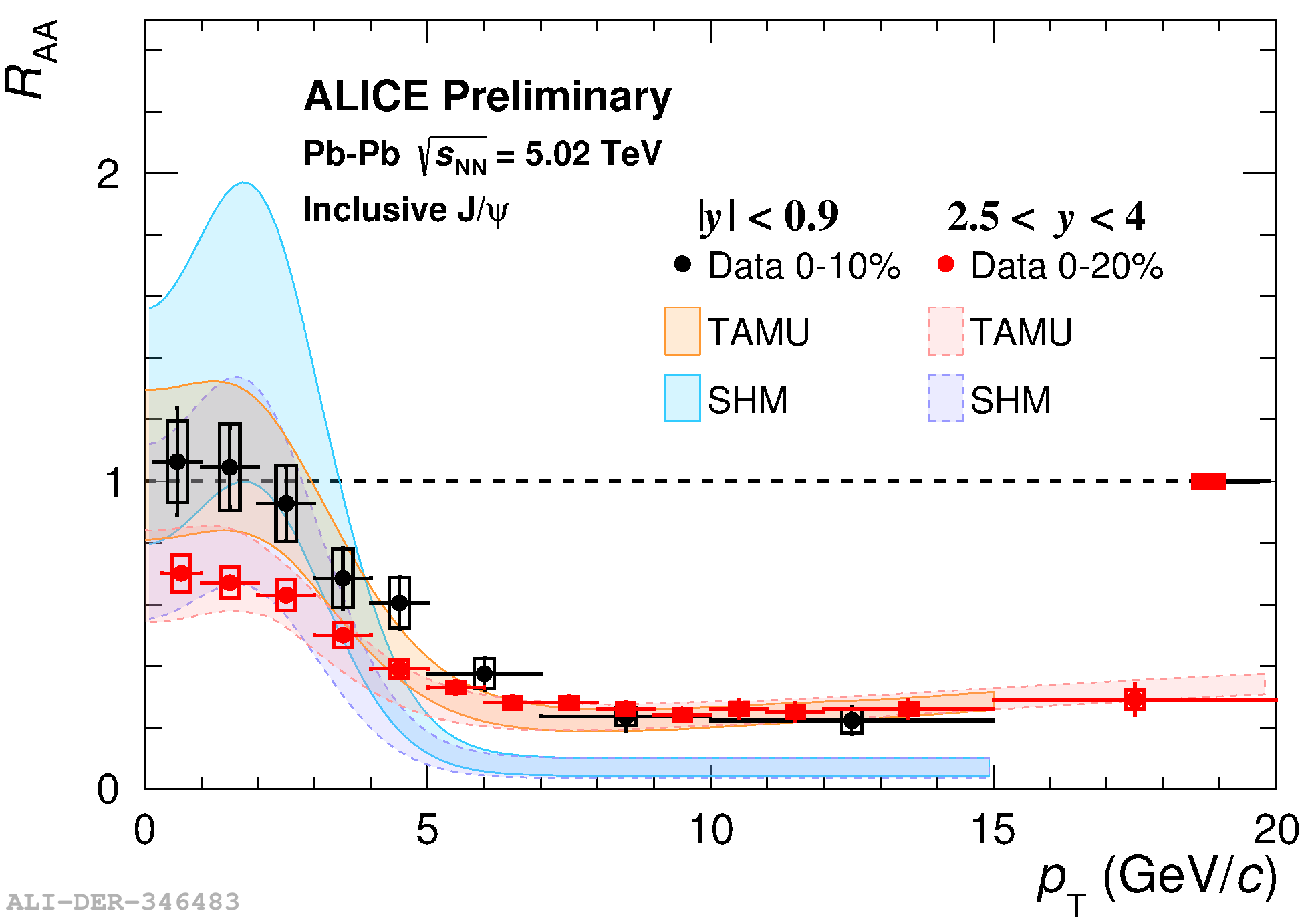}
\caption{Preliminary measurement of the \jpsi nuclear modification factor in central \PbPb collisions at \fivenn at forward and central rapidity in red and black. The measurements are compared to statistical hadronisation and transport model calculations~\cite{SHM,TAMU}. }
\label{jpsRaa}       % Give a unique label
\end{figure}

\begin{figure}[h]
\centering
\includegraphics[width=8.5cm,clip]{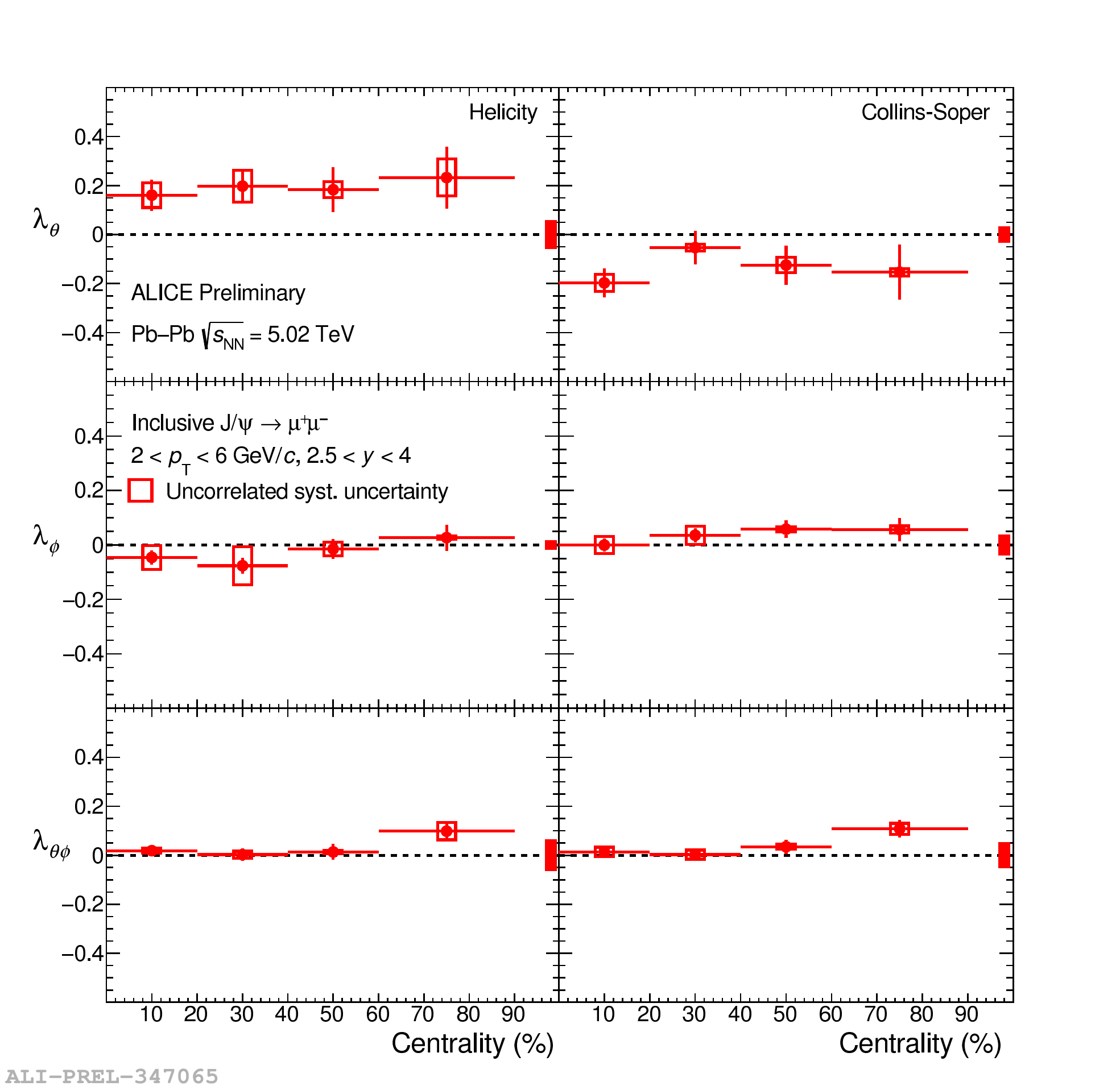}
\caption{\jpsi polarization parameters as a function of centrality measured at forward rapidity in \PbPb collisions at \fivenn.}
\label{polarization}       % Give a unique label
\end{figure}

\indent The medium effect on charmonium production is quantified using the nuclear modification factor (\Raa) calculated as the ratio between the measured yield in \PbPb and \pp collisions at the same energy scaled by the number of binary collisions.
%\begin{equation}
%\Raa = \frac{\mathrm{d}N_\mathrm{AA}/\mathrm{d}\pt}{\langle \Ncoll \rangle \times \mathrm{d}N_\mathrm{pp}/\mathrm{d}\pt} \,
%\end{equation}
%where $\langle \Ncoll \rangle$ is the number of binary collisions, and $\mathrm{d}N_\mathrm{AA}/\mathrm{d}\pt $ and $\mathrm{d}N_\mathrm{pp}/\mathrm{d}\pt $ are the measured yields in \PbPb and \pp collisions, respectively. 
Figure~\ref{jpsRaa} shows the preliminary \Raa in central \PbPb collisions at \fivenn, measured at both forward and central rapidity as a function of \pt. These measurements are obtained from a significantly larger data sample than previous measurements at the same collisions energy, allowing for a more precise and differential result~\cite{ALICE:midRap, ALICE:forwardRap}.  In the low \pt region (< 5 \GeVc) the \jpsi mesons show less suppression at midrapidity than at forward rapidity. This is consistent with the \jpsi (re)generation scenario, as a higher charm cross section is expected at midrapidity. The measurements are compared to statistical hadronisation model (SHM)~\cite{SHM} calculations, where the production of bound states is determined by statistical weights at the phase boundary, and transport model (TAMU)~\cite{TAMU} calculations assuming continuous production and dissociation of charmonium states throughout the whole QGP phase. Within the current uncertainties both models describe the observed trend.
\\
\\
\indent Another physical observable is the quarkonium polarization, measured through the anisotropies in the angular distribution of the decay products, 
\begin{equation}
					W(\cos \theta, \phi ) \propto \frac{1}{3 + \lambda_\theta} \cdot (1+\lambda_\theta \cos^2\theta + 
\lambda_\phi \sin^2\theta\cos2\phi +
\lambda_{\theta\phi} \sin2\theta\cos\phi ) \, ,
\nonumber
\end{equation}
where $\theta$ and $\phi$ are the polar and azimuthal angles in the adopted reference frame.
 %The left panel of Fig.~\ref{polarization} shows the first measurement of \jpsi polarization in \PbPb collisions, measured at forward rapidity as a function of \pt~\cite{ALICE:polarization}. The polarization parameters are shown both in the helicity and collins-soper reference frame, and have values close to zero. For the $ \lambda_\theta$ a maximum of $2\sigma$ deviation with respect to zero is observed for $2 < \pt < 4$~\GeVc. The measured polarization is compatible with ALICE measurements in \pp collisions~\cite{ALICE:polarizationPP}, while a $3\sigma$ difference is observed with \pp results from LHCb in the helicity reference frame~\cite{LHCb:polarization}. This might be reflecting the different production and suppression mechanisms in \PbPb with respect to \pp collisions. 
The first measurement of \jpsi polarization in \PbPb collisions at \fivenn as a function of \pt was recently reported by ALICE~\cite{ALICE:polarization}, showing a $\lambda_\theta$  with a maximum of $2\sigma$ deviation with respect to zero in the \pt interval from 2 -- 4~\GeVc. The preliminary measurement of the \jpsi polarization in \PbPb collisions at \fivenn as a function of centrality is shown in Fig.~\ref{polarization}. The polarization parameters show a flat trend in both the helicity and Collins-Soper reference frames. A non-zero $\lambda_\theta$ is observed in both frames, with slightly positive values in the helicity frame and slightly negative values in the Collins-Soper frame. 

%\begin{figure}[h]
%\centering
%\includegraphics[width=7.1cm,clip]{figures/polarizationVsPt}
%\includegraphics[width=7.cm,clip]{figures/polarizationVsCentrality}
%\caption{\jpsi polarization parameters as a function of \pt and centrality}
%\label{polarization}       % Give a unique label
%\end{figure} 

%\\
%\\
%\indent The \jpsi nuclear modification factor as a function of \Npart is shown in the left panel of Fig.~\ref{jpsiExcess}  where an increase in the \Raa at very low \pt (< 0.3 \GeVc) is observed towards more peripheral collisions. The \Raa in the interval $0-0.3$~\GeVc is compared to a \pt reference interval between 1 and 2~\GeVc and shows a systematically larger \Raa than that of the reference interval. The reference interval is chosen to be in a \pt range where only hadronic processes are expected to contribute to the measured \jpsi yield.

%\begin{figure}[h]
%\centering
%\includegraphics[width=7.cm,clip]{figures/Raa_centrality_excess_1.pdf}
%\includegraphics[width=7.cm,clip]{figures/CrossSection_centrality_models_0.pdf}
%\caption{Left: \jpsi nuclear modification factor at forward rapidity as a function of $\langle \Npart\rangle$ for three \pt intervals. Right: \jpsi coherent photoproduction cross section in \PbPb collisions at \s $= 5$ and $2.76$ \TeV~\cite{ALICE:excessYield}.}
%\label{jpsiExcess}       % Give a unique label
%\end{figure}

%\indent The underlying physics mechanism suggested to cause the increasing \Raa in peripheral collisions at very low \pt is coherent \jpsi photoproduction. Based on this assumption, the coherent \jpsi photoproduction cross section is extracted from the \jpsi excess yield by correcting for the fraction of \jpsi from incoherent photoproduction and coherent photoproduction \jpsi decaying from \psiprime. 

\begin{figure}[h]
\centering
\includegraphics[width=8.5cm,clip]{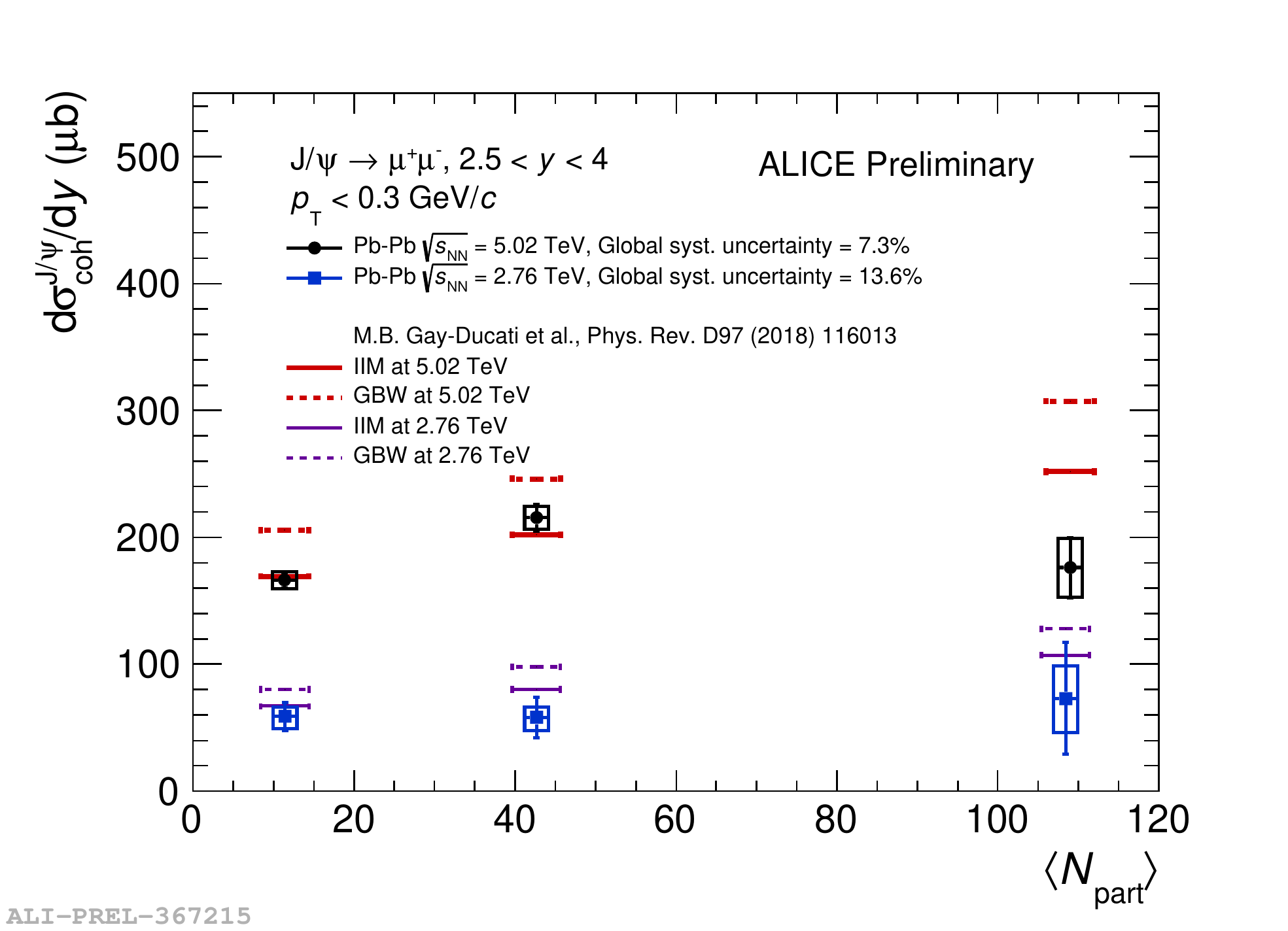}
\caption{\jpsi coherent photoproduction cross section measured at forward rapidity in \PbPb collisions at \snn $= 5$ and $2.76$ \TeV~\cite{ALICE:excessYield} in black and blue, respectively.}
\label{jpsiExcess}       % Give a unique label
\end{figure}

\indent The preliminary measurement of the coherent \jpsi photoproduction cross section at \fivenn measured at forward rapidity is shown in Fig.~\ref{jpsiExcess}. The measured coherent \jpsi photoproduction cross section shows an increase by a factor $\sim3$ with respect to the coherently photoproduced \jpsi at \twosevensixnn~\cite{ALICE:excessYield}. The coherent photoproduction cross section is compared with models implementing a modification of the photon flux with respect to ultra-peripheral collisions (UPC). A light cone color dipole formalism was used in the Golec-Biernat-Wusthoff (GBW) calculation, while the Iancu-Itakura-Munier (IIM) calculation is based on the Color Glass Condensate approach~\cite{upcModel}. A qualitative agreement is observed for the most peripheral collisions, while the measurements in semicentral events deviate from the model predictions.
\\
\\
\indent In summary, ALICE provides a comprehensive set of quarkonium results in heavy-ion collisions. The inclusive \jpsi \Raa measured at both forward and central rapidity as a function of transverse momentum is consistent with a significant contribution from (re)generation in the low \pt region. The \jpsi polarization is measured for the first time in \PbPb collisions, showing a flat trend for all polarization parameters as a function of centrality, with a non-zero $\lambda_\theta$. Measurements of the coherent \jpsi photoproduction cross section at forward rapidity show a considerable increase in the cross section for increasing center-of-mass energy. The photoproduction cross section is well described by UPC-based models in peripheral collisions, but deviations are observed for semicentral collisions.

\end{document}